\documentclass[prl,twocolumn,showpacs,amsmath,amssymb]{revtex4-1}
\usepackage{graphicx}
\usepackage{enumerate}
\usepackage{color}
\usepackage[normalem]{ulem}  % \sout{old text} for strikeout

\renewcommand\sout{\bgroup \color{red} \ULdepth=-.5ex \ULset}

\begin{document}

\title{Fate of the Tetraquark Candidate $Z_{c}(3900)$ from Lattice QCD}

\author{
Yoichi~Ikeda$^{1,2}$,
Sinya~Aoki$^{3,4}$,
Takumi~Doi$^{2}$,
Shinya Gongyo$^{3}$,
Tetsuo~Hatsuda$^{2,5}$,
Takashi~Inoue$^{6}$,
Takumi Iritani$^{7}$,
Noriyoshi~Ishii$^{1}$,
Keiko~Murano$^{1}$,
Kenji~Sasaki$^{3,4}$\\ (HAL QCD Collaboration)
}

\affiliation{
$^1${Research Center for Nuclear Physics (RCNP), Osaka University, Osaka 567-0047, Japan}\\
$^2${Theoretical Research Division, Nishina Center, RIKEN, Saitama 351-0198, Japan}\\
$^3${Yukawa Institute for Theoretical Physics, Kyoto University, Kyoto 606-8502, Japan}\\
$^4${Center for Computational Sciences, University of Tsukuba, Ibaraki 305-8571, Japan}\\
$^5${iTHES Research Group, RIKEN, Saitama 351-0198, Japan}\\
$^6${Nihon University, College of Bioresource Sciences, Kanagawa 252-0880, Japan}\\
$^7${Department of Physics and Astronomy, Stony Brook University, New York 11794-3800, USA}
}

\begin{abstract}
The possible exotic meson $Z_{c}(3900)$, found in $e^+ e^-$ reactions, is studied by the method of coupled-channel scattering in lattice QCD.
The interactions among $\pi J/\psi$, $\rho \eta_{c}$ and $\bar{D}D^{*}$ channels are derived from (2+1)-flavor QCD simulations at $m_{\pi}=410 \text{--} 700$ MeV.
The interactions are dominated by the off-diagonal $\pi J/\psi$-$\bar{D}D^{*}$ and $\rho \eta_{c}$-$\bar{D}D^{*}$ couplings, which indicates that the $Z_{c}(3900)$ is not a usual resonance but a threshold cusp. 
Semiphenomenological analyses with the coupled-channel interaction are also presented to confirm this conclusion.
\end{abstract}
%% 12.38.Gc : LQCD calculations
%% 14.40.Rt : exotic mesons
%% 13.75.Lb : Meson-meson interactions
%% 13.25.Gv : Decays of J/psi, Y, and other quarkonia
%% 11.80.Gw : Multichannel scattering

\maketitle

%%%%%%%%%%%%%%%%%%%%%%%%%%%%%%%%%

\begin{figure}[htb]
\includegraphics[width=0.45\textwidth,clip]{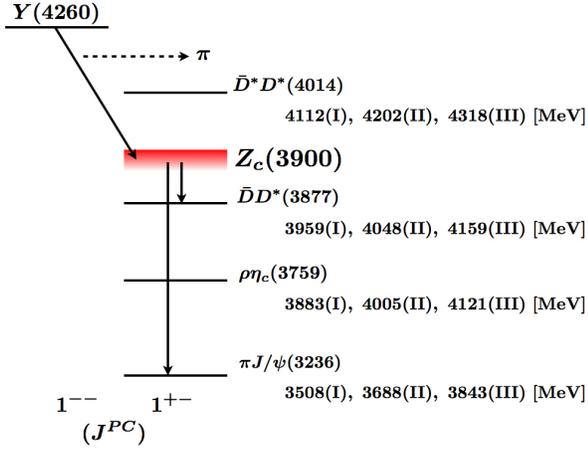}
\caption{
A possible decay scheme of the $Y(4260)$ through $Zc(3900)$, 
together with the relevant two-meson thresholds of the $Z_c(3900)$ decay 
at $m_{\pi}\simeq 140$ (Expt.), $410$ (case I), $570$ (case II) and $700$ (case III) MeV.
The arrows represent the observed decay modes in the experiments~\cite{expt_BESIII,expt_Belle,expt_CLEO-c}.
}
\label{fig1}
\end{figure}

%%%%%%%%%%%%%%%%%%%%%%%%%%%%%%%%%%%%%%%%%
%   Introduction
%%%%%%%%%%%%%%%%%%%%%%%%%%%%%%%%%%%%%%%%%%%%%%%%%%%%%%%%%%%%%%%%%%%%%%%%%%
One of the long-standing problems in hadron physics is to identify the existence of exotic hadrons 
different from the quark-antiquark states (mesons) and three-quark states (baryons).
Candidates of such exotic hadrons include 
the pentaquark states $P_c^+(4380)$ and $P_c^+(4450)$ 
observed by the LHCb Collaboration~\cite{Aaij:2015tga}
and the tetraquark states $Z_c(3900)$ 
reported by the BESIII~\cite{expt_BESIII}, the Belle~\cite{expt_Belle}, 
and the CLEO-c \cite{expt_CLEO-c} Collaborations.
In particular, $Z_c(3900)$ appears as a peak 
in both the $\pi^{\pm} J/\psi$ and $\bar{D} D^{*}$ invariant mass spectra in the reaction, 
$e^{+} e^{-} \to Y(4260) \to \pi^{\pm} \pi^{\mp} J/\psi,  \pi \bar{D} D^{*}$: 
Its quantum numbers are then identified as $I^G (J^{PC})=1^+(1^{+-})$, 
so that at least four quarks, $c\bar{c}u\bar{d}$ (or its isospin partners), are involved. 
(See the level structure and the decay scheme in Fig.\ref{fig1}.)

So far, there have been various phenomenological attempts 
to characterize the $Z_c(3900)$ as a hadro-charmonium, a compact tetraquark, a hadronic molecule 
(e.g., Refs.~\cite{model_Voloshin, model_Cleven}) 
as well as a kinematical threshold effect (e.g., Refs.~\cite{model_Matsuki,model_Swanson}).
However, due to the lack of information of the diagonal and off-diagonal interactions 
among different channels (such as $\pi J/\psi$, $\rho \eta_c$, and $\bar{D}D^{*}$), 
the predictions of those models are not well under theoretical control.
On the other hand, the direct lattice QCD studies with the standard method of temporal correlations 
show no candidate for the  $Z_c(3900)$ eigenstate~\cite{LQCD_Sasa, LQCD_Lee}, 
which indicates that the $Z_c(3900)$ may not be an ordinary resonance state.
Under these circumstances, it is most desirable to carry out manifestly coupled-channel analyses 
with the first-principles QCD inputs.
%%%%

The purpose of this Letter is to report a first attempt to determine the nature of the $Z_c(3900)$ 
on the basis of the HAL QCD method~\cite{Ishii2007a,Aoki2010,Ishii2012,HAL2012}.   
We consider three two-body channels below $Z_{c}(3900)$
($\pi J/\psi$, $\rho \eta_c$ and $\bar{D}D^{*}$) which couple with each other. 
The interactions among these channels faithful to the QCD $S$ matrix are derived 
from the equal-time Nambu-Bethe-Salpeter (NBS) wave functions on the lattice 
according to the coupled-channel formulation of the HAL QCD method~\cite{Aoki2011,Aoki2012,Sasaki2015}. 
The $s$-wave interactions and the $S$ matrix thus obtained 
are used to search for the complex poles in the $\pi J/\psi$ and $\bar{D}D^{*}$ scattering amplitudes 
to unravel the nature of the $Z_c(3900)$.
We note here that the conventional resonances such as the $\rho$ meson and the $\Delta$ baryon 
have not yet been analyzed in the HAL QCD method, 
and the comparison with the L\"{u}scher's method~\cite{Luscher:1990ux} in these channels 
is one of the important future subjects.
(Such comparison in the nonresonant $\pi\pi$ channel has been done 
in Refs.~\cite{Kurth:2013tua, Iritani:2015dhu}; See also Ref.~\cite{Iritani:2016jie}).
Note also that the coupled-channel HAL QCD method has not been experimentally tested in other systems yet:
Baryon-baryon interactions with hyperons are currently underway~\cite{Sasaki:2016gpc} for comparison with future experimental data.
With these reservations in mind, we extract invariant mass spectra of the three-body decays  $Y(4260) \rightarrow \pi \pi J/\psi$ and $\pi \bar{D} D^{*}$ using the scattering amplitudes obtained in lattice QCD, and the results are then compared with experimental data.

%%%%%%%%%%%%%%%%%%%%%%%%%%%%%%%%%%%%
%%%%%%%%%%%%%%%%%%%%%%%%%%%%%%%%%%%%%%%%%%%%%%%%%

%%%%%%%%%%%%%%%%%%%%%%%%%%%%%%%%%%%%%%%%%%%%%%%%%%%%%%%%%%%%%%%%%%%%%%%%%%
%   HAL QCD approach for coupled-channel scatterings
%%%%%%%%%%%%%%%%%%%%%%%%%%%%%%%%%%%%%%%%%%%%%%%%%%%%%%%%%%%%%%%%%%%%%%%%%%
The starting point of the coupled-channel HAL QCD method \cite{Aoki2011,Aoki2012,Sasaki2015} is a normalized correlation function
%--------------------------------------------------------
\begin{eqnarray}
 C^{\alpha \beta}(\vec{r},t) \equiv 
 \sum_{\vec{x}} \left\langle 0 \right| \phi^{\alpha}_{1}(\vec{x} + \vec{r}, t) \phi^{\alpha}_{2}(\vec{x}, t) 
 \overline{\mathcal{J}}^{\beta} \left| 0 \right\rangle /{\sqrt{Z_{1}^{\alpha} Z_{2}^{\alpha}}}, \nonumber \\
\label{eq:Cab}
\end{eqnarray}
%--------------------------------------------------------
where each channel is specified by $\alpha=(\pi J/\psi, \rho \eta_c, \bar{D}D^{*})$,
and $\phi^{\alpha}_{i}(\vec{y},t)$ is a local Heisenberg operator at Euclidian time $t >0$ and the spatial point $\vec{y}$ for the meson $i \ (=1,2)$ with mass $m_{i}^{\alpha}$ in channel $\alpha$.
The corresponding wave function renormalization factor is given by $Z_{i}^{\alpha}$.
$\overline{\mathcal{J}}^{\beta}$ denotes a two-meson operator in channel $\beta$ with zero-momentum wall quark source located at $t=0$.
The NBS wave function $\psi^{\alpha}_{n}(\vec{r})$ for each scattering state specified by $n$ on the lattice is related to Eq.(\ref{eq:Cab}) as $C^{\alpha \beta}(\vec{r},t)= \sum_n  \psi^{\alpha}_{n}(\vec{r}) A_{n}^{\beta} e^{-W_n t}$ with $W_n$ being the  eigenvalue of the $n$ th QCD eigenstate.
$A_{n}^{\beta} \equiv \langle W_n | \overline{\mathcal{J}}^{\beta} | 0 \rangle$ is an overlap between the eigenstate and QCD vacuum by the insertion of $\overline{\mathcal{J}}^{\beta}$.
It can be shown that $R^{\alpha \beta}(\vec{r},t) \equiv C^{\alpha \beta}(\vec{r},t) e^{(m_1^\alpha+m_2^\alpha)t}$
satisfies the Schr\"{o}dinger-type equation~\cite{Ishii2012,Aoki2012},
%-----------------------------------------------------------------------
\begin{eqnarray}
&& 
\biggl( -\frac{\partial}{\partial t} - H^{\alpha}_{0} \biggr)
R^{\alpha \beta}(\vec{r}, t) 
=  
\nonumber \\
& & ~~~~~~~~~~~~~~
\sum_{\gamma} \Delta^{\alpha \gamma}
\int d\vec{r'}  U^{\alpha \gamma}(\vec{r},\vec{r'}) 
R^{\gamma \beta}(\vec{r'}, t) ~ ,
\label{t-dep_local}
\end{eqnarray}
%-----------------------------------------------------------------------
where $H^{\alpha}_{0}=-\nabla^{2}/2\mu^{\alpha}$ with the reduced mass $\mu^{\alpha}=m^{\alpha}_{1} m^{\alpha}_{2} /(m^{\alpha}_{1} + m^{\alpha}_{2})$
and $\Delta^{\alpha \gamma}=e^{(m_{1}^{\alpha}+m_{2}^{\alpha}) t}/e^{(m_{1}^{\gamma}+m_{2}^{\gamma}) t}$.
In the above equation, we neglect terms associated with relativistic corrections, 
${\mathcal O}\bigl( (\partial_{t}^{2} / m^{\alpha}_{1,2}) (\partial_{t} / m^{\alpha}_{1,2})^{n} \bigr)$
with $n \ge 0$. 
We have checked that the relativistic corrections are negligible in the present lattice setup with relatively large pion masses.
Here we consider $t$ sufficiently large so that the inelastic states (The lowest one is $\bar{D}^{*}D^{*}$ in the present lattice QCD setup) becomes negligible in $U^{\alpha \beta}$, otherwise these channels should be taken into account explicitly.
The energy-independent coupled-channel potential $U^{\alpha \beta}(\vec{r},\vec{r'})$ guarantees that 
the $S$ matrix is unitary below the $\bar{D}^{*}D^{*}$ threshold~\cite{Aoki2011,Aoki2012}
and gives the correct scattering amplitude. 
In the following, we take the $s$-wave projection ($A^{+}_{1}$ projection of the cubic group on the lattice) and also employ the lowest order of the velocity expansion, 
$U^{\alpha \beta}(\vec{r},\vec{r'})= V^{\alpha \beta}(\vec{r}) \delta(\vec{r}-\vec{r'})+O(\nabla^2)$ 
to extract the spherical and local potential $V^{\alpha \beta}( r )$. 
The systematic errors originating from higher derivative terms are 
estimated by the $t$ dependence of the observables~\cite{Ishii2012}.

We note here that the HAL QCD method and the conventional L\"{u}scher's method are both based on Eq.~(\ref{eq:Cab}).
In the coupled-channel L\"{u}scher's method proposed in Refs.~\cite{Doring:2011vk, Wu:2014vma, Dudek:2014qha},
some phenomenological parametrization of the $K$ matrix is employed 
that approximates the energy dependence of the coupled-channel $S$ matrix, 
while in the present method, the velocity expansion is employed that approximates the nonlocality of the coupled-channel potentials.

%%%%%%%%%%%%%%%%%%%%%%%%%%%%%%%%%%%%%%%%%%%%%%%%%%%%%%%%%%%%%%%%%%%%%%%%%%
%  Numerical setup of LQCD simulations
%%%%%%%%%%%%%%%%%%%%%%%%%%%%%%%%%%%%%%%%%%%%%%%%%%%%%%%%%%%%%%%%%%%%%%%%%%
In order to extract $V^{\alpha \beta}( r )$ from lattice QCD simulation,
we employ (2+1)-flavor QCD gauge configurations generated by the PACS-CS Collaboration~\cite{PACS-CS2009,PACS-CS2010} on a $32^3 \times 64$ lattice with the renormalization group improved gauge action at $\beta_{\rm lat} = 1.90$ and the nonperturbatively $O(a)$-improved Wilson quark action at $C_{\rm SW}=1.715$.
These parameters correspond to the lattice spacing $a = 0.0907(13)$ fm and the spatial lattice volume $L^3 \simeq (2.9 ~ {\rm fm})^3$.
The hopping parameters are taken to be $\kappa_{ud} = \text{0.13 700, 0.13 727, 0.13 754}$ for $u$ and $d$ quarks and $\kappa_s = \text{0.13 640}$ for the $s$ quark. 
We employ the relativistic heavy quark action for the charm quark~\cite{RHQ_Aoki} to remove the leading order and next-to-leading order cutoff errors, ${\mathcal O}((m_c a)^{n})$ and ${\mathcal O}((m_c a)^{n} (a \Lambda_{\rm QCD}))$, respectively~\cite{Namekawa2011,Ikeda2013}.
To improve the statistics, measurements are repeated twice for each configuration by shifting the source in time direction.
The statistical errors are evaluated by the jackknife method.
The calculated meson masses and the number of configurations $N_{\mathrm{cfg}}$ used in our simulations are listed in Table~\ref{tab1} together with the physical meson masses.
The two-meson thresholds relevant to our analysis are shown in Fig.~\ref{fig1}: 
Because of the heavy pion mass in our simulation, the $\pi \psi'(3826)$ threshold is above the $\bar{D}D^{*}$ threshold.
Also, $\rho \to \pi \pi$ decay is not allowed with $L \simeq 3$fm, so that $\rho \eta_c$ is a well-defined two-body channel.
Pair annihilations of charm quarks are not considered in the present simulations.
%%%%%%%%%%%%%%%%%%%%%%%%%%%%%%%%%%%%%%%%%%%%%%%%%%%%%%%%%%%%%%%%%%%%%%%%%%
\begin{table}[tbhp]
   \centering
   %\topcaption{Table captions are better up top} % requires the topcapt package
   \begin{tabular}{c|ccccccc}
      \hline
      \hline
         & $m_{\pi}$ & $m_{\rho}$ & $m_{\eta_c}$ & $m_{J/\psi}$ & $m_{\bar{D}}$ & $m_{D^{*}}$ & $N_{\mathrm{cfg}}$   \\
      \hline 
Expt.    & 140   &  775    & 2984    & 3097    & 1870    & 2007    & \\
      \hline 
Case I   & 411(1) &  896(8) & 2988(1) & 3097(1) & 1903(1) & 2056(3) & 450 \\
      \hline 
Case II  & 570(1) & 1000(5) & 3005(1) & 3118(1) & 1947(1) & 2101(2) & 400 \\
      \hline 
Case III & 701(1) & 1097(4) & 3024(1) & 3143(1) & 2000(1) & 2159(2) & 399 \\
      \hline 
      \hline 
   \end{tabular}
   \caption{ Meson masses in MeV units and the number of configurations used 
   in our simulations. }
   \label{tab1}
\end{table}
%%%%%%%%%%%%%%%%%%%%%%%%%%%%%%%%%%%%%%%%%%%%%%%%%%%%%%%%%%%%%%%%%%%%%%%%%%

%%%%%%%%%%%%%%%%%%%%%%%%%%%%%%%%%%%%%%%%%%%%%%%%%%%%%%%%%%%%%%%%%%%%%%%%%%
%\section{Coupled-channel potentials and structure of the $Z_c(3900)$ in $I^{G}(J^{PC})=1^+(1^{+-})$}
%%%%%%%%%%%%%%%%%%%%%%%%%%%%%%%%%%%%%%%%%%%%%%%%%%%%%%%%%%%%%%%%%%%%%%%%%%

%%%%%%%%%%%%%%%%%%%%%%%%%%%%%%%%%%%%
\begin{figure*}[!t]
\includegraphics[width=0.95\textwidth,clip]{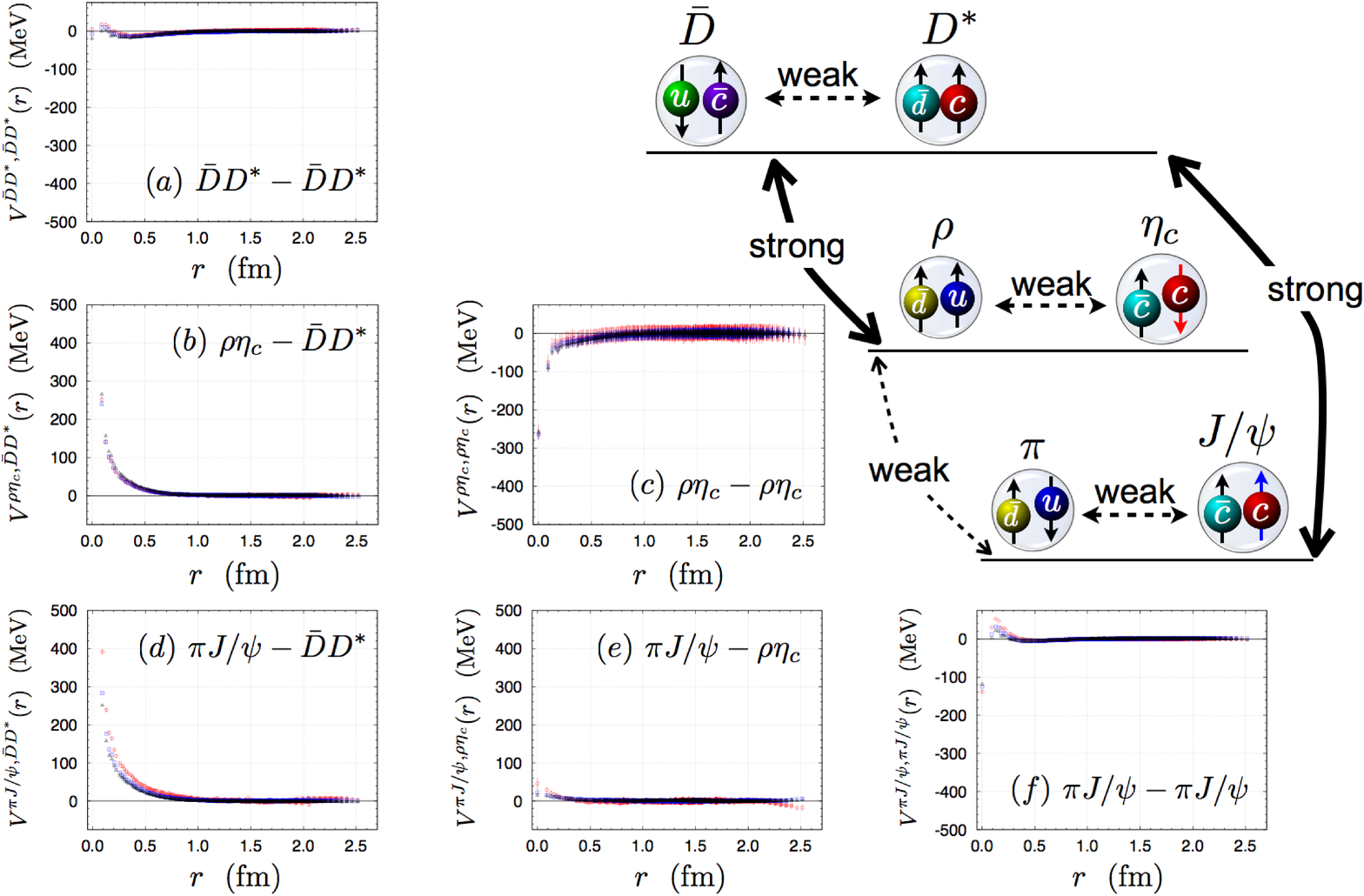}
\caption{
The $s$-wave potentials for the (a) $\bar{D}D^{*}$-$\bar{D}D^{*}$, (b) $\rho \eta_c$-$\bar{D}D^{*}$, (c) $\rho \eta_c$-$\rho \eta_c$, (d) $\pi J/\psi$-$\bar{D}D^{*}$, (e) $\pi J/\psi$-$\rho \eta_c$ and (f) $\pi J/\psi$-$\pi J/\psi$ channels.
The coupled-channel potentials are obtained at time slice $t=13$ for 
case I(red circles), case II(blue squares) and case III(black triangles).
}
\label{fig2}
\end{figure*}
%%%%%%%%%%%%%%%%%%%%%%%%%%%%%%%%%%%%%%%%%%%%%%%%%
In Fig.~\ref{fig2}, we show the results of the $s$-wave $\pi J/\psi$-$\rho \eta_c$-$\bar{D}D^{*}$ coupled-channel potentials at time slice $t=13$,
where the time-slice dependence in $t=11\text{--}15$ on the potentials $V^{\alpha \beta}$ is found to be weak:
This implies that contributions from the inelastic $\bar{D}^{*}D^{*}$ scattering states to $V^{\alpha \beta}$ are negligible, 
and the convergence of the derivative expansion is reliable.
We find that all diagonal potentials, 
(a) $V^{\bar{D}D^{*}, \bar{D}D^{*}}$, (c) $V^{\rho \eta_c, \rho \eta_c}$, and (f) $V^{\pi J/\psi, \pi J/\psi}$ are very weak.
This observation indicates that the $Z_c(3900)$ is neither a simple $\pi J/\psi$ nor $\bar{D}D^{*}$ molecule.
Among the off-diagonal potentials, we find that the $\pi J/\psi$-$\rho \eta_c$ coupling in Fig.~\ref{fig2} (e) is also weak:
This is consistent with the heavy-quark spin symmetry, 
which tells us that the spin flip amplitudes of the charm quark are suppressed by ${\cal O}(1/m_c)$.
On the other hand, (b) the $\rho \eta_c$-$\bar{D}D^{*}$ coupling and (d) the $\pi J/\psi$-$\bar{D}D^{*}$ coupling are both strong: 
They correspond to the rearrangement of quarks between the hidden charm sector and the open charm sector.

%%%%%%%%%%%%%%%%%%%%%%%%%%%%%%%%%%%%
%%  Two-body observables
%%%%%%%%%%%%%%%%%%%%%%%%%%%%%%%%%%%%

%%%%%%%%%%%%%%%%%%%%%%%%%%%%%%%%%%%%%
\begin{figure*}[!t]
\begin{center}
\includegraphics[width=0.85\textwidth,clip]{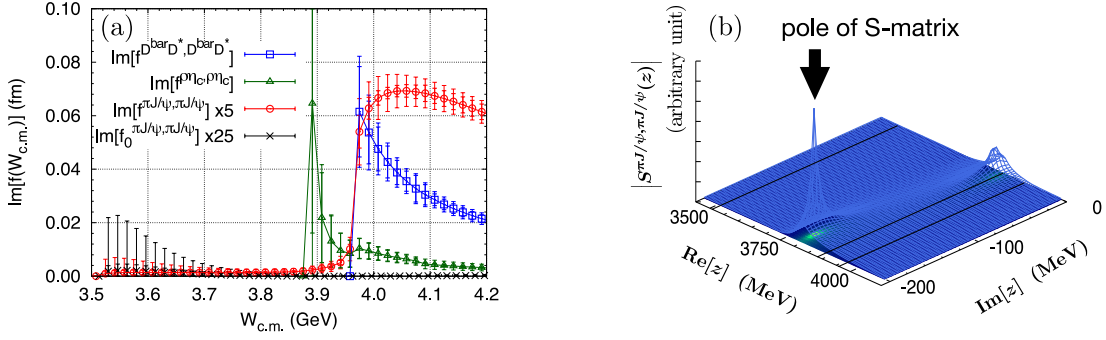}
\end{center}
\caption{
(a) The two-body invariant mass spectra in the $\pi J/\psi$ (red circles, scaled by 5), $\rho \eta_{c}$ (green triangles) and $\bar{D}D^{*}$ (blue squares) channels.
The two-body $\pi J/\psi$ spectrum without the off-diagonal component of $V^{\alpha \beta}$ is also shown
by $\mathrm{Im} f_{0}^{\pi J/\psi, \pi J/\psi}$ (black crosses, scaled by 25).
(b) The pole of the $S$ matrix on the [$bbb$] sheet in the notation of Ref.~\cite{Pearce:1988rk} for $\pi J/\psi$, $\rho \eta_c$ and $\bar{D}D^{*}$ channels 
($z = m_1^{\alpha} + m_2^{\alpha} + p_{\alpha}^2/2\mu^{\alpha}$).
Both figures correspond to the case I in Table \ref{tab1}.
In Fig.~\ref{fig3}(a), 
the inner error is statistical, while the outer one is statistical and systematic combined in quadrature.
}
\label{fig3}
\end{figure*}
%%%%%%%%%%%%%%%%%%%%%%%%%%%%%%%%%%%%%%%%%%%%%%%%%%%%%%%%%%%%%%%%%%%%%%%%%%

As a first step to investigate the structure of the $Z_c(3900)$ on the basis of $V^{\alpha \beta}$ just obtained, 
let us consider the two-body $T$ matrix [see, e.g., Eq. (16.43) of Ref.~\cite{Newton:book}]:
%-------------------------------------------------------------------------
\begin{eqnarray}
&  &
t^{\alpha \beta}(\vec{p}_{\alpha}, \vec{p}_{\beta}; W_{\mathrm{c.m.}})
=  
V^{\alpha \beta}(\vec{p}_{\alpha}, \vec{p}_{\beta})  \nonumber \\
&  & ~~~~
+ \sum_{\gamma} \int d\vec{q_{\gamma}}
\frac{ V^{\alpha \gamma}(\vec{p}_{\alpha}, \vec{q}_{\gamma}) 
t^{\gamma \beta}(\vec{q}_{\gamma}, \vec{p}_{\beta}; W_{\mathrm{c.m.}})
}{
W_{\mathrm{c.m.}} - E_{\gamma}(\vec{q}_{\gamma}) + i \epsilon
} ~,
\end{eqnarray}
%-------------------------------------------------------------------------
where $\vec{p}_{\alpha}$ ($\vec{q}_{\gamma}$) indicates the on-shell (off-shell) momentum of the two-meson state in channel $\alpha$ ($\gamma$).
$W_{\mathrm{c.m.}}$ and $E_{\gamma}(\vec{q}_{\gamma})$ represent the scattering energy in the center-of-mass (c.m.) frame and the energy of the intermediate states in channel $\gamma$, respectively.

Shown in Fig.~\ref{fig3} (a) are the invariant mass spectra 
${\rm Im} f^{\alpha \alpha}(W_{\rm c.m.}) = - \pi \mu^{\alpha} {\rm Im} t^{\alpha \alpha}(W_{\rm c.m.})$
in the $\pi J/\psi$ (red circles), $\rho \eta_c$ (green triangles) and $\bar{D}D^{*}$ (blue squares) channels obtained from lattice QCD for case I in Table \ref{tab1}.
The amplitude $f^{\alpha \beta}(W_{\rm c.m.})$ is related to the differential cross section as 
$d\sigma^{\alpha \beta}/d\Omega = |f^{\alpha \beta}(W_{\rm c.m.})|^2$.
In Fig.~\ref{fig3} (a), the inner errors are statistical only, while the outer ones are statistical and systematic errors added in quadrature:
The systematic errors from the truncation of the derivative expansion are evaluated by the difference between ${\rm Im} f^{\alpha \alpha}$ at $t=13$ and that at $t=15$.
The peak structures in $\rho \eta_c$ and $\bar{D}D^{*}$ spectra are caused by the opening of the $s$-wave thresholds.
The sudden enhancement in the $\pi J/\psi$ spectrum just above the $\bar{D}D^{*}$ threshold is induced by the $\pi J/\psi$-$\bar{D}D^{*}$ coupling.
Indeed, if we switch off the off-diagonal components of $V^{\alpha \beta}$, 
the red circles turn into the black crosses without any peak structure.
This implies that the peak structure in the $\pi J/\psi$ spectrum [called $Z_c(3900)$]
is a typical ``threshold cusp''\cite{Wigner:1948zz,Newton:book} due to the opening of the $s$-wave $\bar{D}D^{*}$ threshold.

To make sure that the $Z_c(3900)$ is not associated with the resonance structure, 
we examine the pole positions of the $S$ matrix on the complex energy plane
according to the notation and procedure in Ref.~\cite{Pearce:1988rk}.
The complex energy is defined as 
$z = m_1^{\alpha} + m_2^{\alpha} + p_{\alpha}^2/2\mu^{\alpha}$, 
and the ``top [$t$]"  (``bottom [$b$]") sheet corresponds to 
$0 \le \arg p_{\alpha} < \pi$ ($\pi \le \arg p_{\alpha} < 2\pi$)
for the complex momentum in each channel ($\alpha = \pi J/\psi, \rho\eta_c, \bar{D}D^{*}$).
Among 8 Riemann sheets for the present three-channel scattering, 
the most relevant one is the [$bbb$] sheet in the notation of Ref.~\cite{Pearce:1988rk}.
We find a pole with a large imaginary part on this sheet (see Supplemental Material~\cite{Suppl}):
$z-(m_{\bar{D}}+m_{D^{*}})=-167(94)(27)-i183(46)(19)$ MeV for case I, 
$-128(76)(33)-i157(32)(19)$ MeV for case II, and $-190(56)(42)-i44(27)(27)$ MeV for case III, 
with the first and second parentheses indicating the statistical and systematic errors, respectively. 
Shown in Fig.~\ref{fig3} (b) is the complex pole on the [$bbb$] sheet for case I. 
It is located far from the $\bar{D}D^{*}$ threshold on the real axis, 
so that the amplitude is hardly affected by the pole. 

%%%%%%%%%%%%%%%%%%%%%%%%%%%%%%%%%%%%%%%%%%%%%%%%%%%%%%%%%%%%%%%%%%%%%%%%%%
% Three-body decay of Y(4260)
%%%%%%%%%%%%%%%%%%%%%%%%%%%%%%%%%%%%%%%%%%%%%%%%%%%%%%%%%%%%%%%%%%%%%%%%%%

%%%%%%%%%%%%%%%%%%%%%%%%%%%%%%%%%%%%%%%%%%%%%%%%%%%%%%%%%%%%%%%%%%%%%%%%%%
\begin{figure}[!th]
\begin{center}
\includegraphics[width=0.425\textwidth,clip]{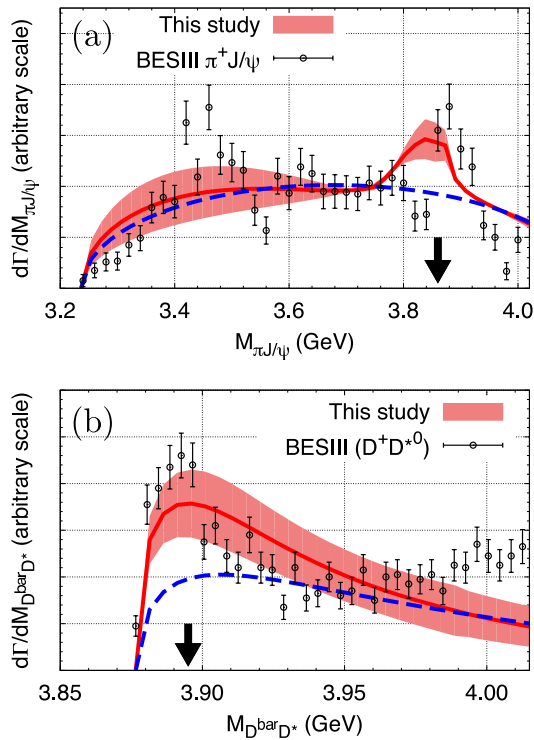}
\end{center}
%\vspace{-0.1cm}
\caption{
The invariant mass spectra of (a) $Y(4260) \to \pi \pi J/\psi$ and (b)  $Y(4260) \to \pi \bar{D}D^{*}$ below the $\bar{D}^{*} D^{*}$ threshold calculated with $V^{\alpha \beta}$ for case I in Table \ref{tab1}.
The shaded areas show the statistical errors.
The vertical arrows show the predicted peak positions from the calculations.
The blue dashed lines show the invariant mass spectra of the $Y(4260)$ decay without the off-diagonal components of $V^{\alpha \beta}$.
The experimental data are taken from Ref.~\cite{expt_BESIII}.
}
\label{fig4}
\end{figure}
%%%%%%%%%%%%%%%%%%%%%%%%%%%%%%%%%%%%%%%%%%%%%%%%%%%%%%%%%%%%%%%%%%%%%%%%%%
To make further connection between the above result and the experimentally observed structure in $\pi J/\psi$ and $\bar{D} D^{*}$ invariant mass spectra~\cite{expt_BESIII, expt_Belle, expt_CLEO-c},
let us now consider semiphenomenological analysis of the three-body decays 
$Y(4260) \to \pi \pi J/\psi$, $\pi \bar{D}D^{*}$ by taking into account the final state rescattering due to $V^{\alpha \beta}$ extracted from lattice QCD simulations.
We model the primary vertex by complex constants 
$C^{Y \rightarrow \pi + \alpha}$ [$\alpha= (\pi J/\psi, \bar{D} D^{*})$].
Then the three-body $T$ matrix $T^{Y \rightarrow  \pi+\beta}$ [$\beta=(\pi J/\psi, \bar{D} D^{*})$]
is given by
%-----------------------------------------------------------------------
\begin{align}
&T^{Y\rightarrow \pi+\beta}(\vec{p}, \vec{q}_{\beta}; W_3) = 
\sum_{\alpha= \pi J/\psi,  \bar{D} D^{*} } 
C^{Y \rightarrow \pi+\alpha}  \nonumber \\
& ~~~ \times
\biggl( 
\delta_{\alpha \beta} + \int d\vec{q}_{\alpha} 
\frac{ 
t^{\alpha \beta}(\vec{q}_{\alpha}, \vec{q}_{\beta}, \vec{p}; W_3) 
}{ W_3 - E_{\pi}(\vec{p}) - E_{\alpha}(\vec{p},\vec{q}_{\alpha}) + i\epsilon }
\biggr) ~,
\label{decay}
\end{align}
%-----------------------------------------------------------------------
where $W_3$, $E_{\pi}(\vec{p})$ and $E_{\alpha}(\vec{p},\vec{q}_{\alpha})$ represent the energies of the $Y(4260)$, the spectator pion with the momentum $\vec{p}$ and the interacting pairs with the relative momentum $\vec{q}_{\alpha}$ in channel $\alpha$, respectively (see Supplemental Material~\cite{Suppl}). 
The decay rate in the rest frame of $Y(4260)$ is obtained as 
%-----------------------------------------------------------------------
$ d\Gamma^{Y \to \pi + \beta}(W_3)  =  (2 \pi)^4 \delta( W_3 - E_{\pi}(\vec{p}) - E_{\beta}(\vec{p},\vec{q}_{\beta}) ) ~ 
d \vec{p} ~ d \vec{q}_{\beta} ~
| T^{Y \to \pi + \beta}(\vec{p}, \vec{q}_{\beta}; W_3) |^2 $.
%-----------------------------------------------------------------------

In order to have the same phase space as the experiments, we employ the physical hadron masses 
while $t^{\alpha \beta}$ is taken from the lattice data for case I. 
The complex couplings $C^{Y\rightarrow \pi+\alpha}$ are fitted to the BESIII data~\cite{expt_BESIII}.
Since the experimental data are in the arbitrary scale,
we focus only on the line shapes of the invariant mass spectra.
In this case, we have two real parameters, 
$R \equiv | C^{Y \rightarrow \pi (\bar{D} D^{*})} / C^{Y \rightarrow \pi (\pi J/\psi)} |$ and $\theta \equiv \arg ( C^{Y \rightarrow \pi (\bar{D} D^{*})} / C^{Y \rightarrow \pi (\pi J/\psi)} )$, 
and the best fit values are $R=0.95(18)$ and $\theta=-58(44)$ degree.
To compare with the raw data of the experiment,
we find 
$\mathcal{N}_{\pi J/\psi} |C^{Y \to \pi (\pi J/\psi)}|=0.40(22)$ GeV$^{-2}$ 
for the $\pi J/\psi$ and 
$\mathcal{N}_{\bar{D} D^{*}} |C^{Y \to \pi (\pi J/\psi)}|=0.76(42)$ GeV$^{-2}$ 
for the $\bar{D} D^{*}$ invariant mass distributions, 
where $\mathcal{N}_{\alpha}$ are the normalization factors to the raw data.
Resulting decay spectra are shown in Figs.~\ref{fig4}(a) and \ref{fig4}(b) where the shaded bands denote the statistical errors:
We find that the coupled-channel potential $V^{\alpha \beta}$ well reproduces the peak structures 
just above the $\bar{D}D^{*}$ threshold at $3.9$ GeV.
%---
The deviation from the experimental data around $4$ GeV may be attributed to the contributions from the higher partial waves between the spectator pion and interacting pairs or from the $\bar{D}^{*}D^{*}$ states, which are not considered in the present study.
%---
If we turn off the off-diagonal components of $V^{\alpha \beta}$ with the same constants 
$C^{Y \rightarrow \pi+\alpha}$, we obtain the results shown by the blue dashed lines,
where the lines are normalized to the results obtained from the full calculations at $4$ GeV.
The peak structures at $3.9$ GeV disappear in this case.

%%%%%%%%%%%%%%%%%%%%%%%%%%%%%%%%%%%%%%%%%%%%%%%%%%%%%%%%%%%%%%%%%%%%%%%%%%
%\section{Summary}
%%%%%%%%%%%%%%%%%%%%%%%%%%%%%%%%%%%%%%%%%%%%%%%%%%%%%%%%%%%%%%%%%%%%%%%%%%
In summary, we have studied the $\pi J/\psi$-$\rho \eta_c$-$\bar{D}D^{*}$ coupled-channel 
interactions using (2+1)-flavor full QCD gauge configurations
in order to study the structure of the tetraquark candidate $Z_c(3900)$.
Thanks to the HAL QCD method, we obtain the full coupled-channel potential $V^{\alpha \beta}$, 
whose diagonal components are all small, 
so that $Z_c(3900)$ cannot be a simple hadro-charmonium or $\bar{D}D^{*}$ molecular state.

Also, we found a strong off-diagonal transition between  $\pi J/\psi$ and $\bar{D}D^{*}$,
which indicates that the $Z_c(3900)$ can be explained as a threshold cusp.
To confirm this, we calculated the invariant mass spectra and pole positions associated with the coupled-channel two-body $S$ matrix on the basis of  $V^{\alpha \beta}$.
The results indeed support that the peak in the  $\pi J/\psi$ invariant mass spectrum is not associated with a conventional resonance state but is a threshold cusp induced by the strong $\pi J/\psi$-$\bar{D}D^{*}$ coupling. 
To further strengthen our conclusion, we made a semiphenomenological analysis of the three-body decay of the $Y(4260)$, and found that the experimentally observed peak structures just above the $\bar{D}D^{*}$ threshold are well reproduced in the $Y(4260) \to \pi \pi J/\psi$ and the $Y(4260) \to \pi \bar{D}D^{*}$ decays.

To make a definite conclusion on the structure of the $Z_{c}(3900)$ in the real world,
we plan to carry out full QCD simulations near the physical point.
It is also an interesting future problem to study the structure of 
pentaquark candidates $P_c^+(4380)$ and $P_c^+(4450)$ on the basis of the coupled-channel HAL QCD method.

%%%%%%%%%%%%%%%%%%%%%%%%%%%%%%%%%%%%%%%%%%%%%%%%%%%%%%%%%%%%%%%%%%%%%%%%%%
\begin{acknowledgments}
The authors thank ILDG/JLDG~\cite{JLDG} for providing us with full QCD gauge configurations used in this study.
Y.I. is grateful to Doctor C.Z. Yuan for providing us with BESIII experimental data.
Numerical calculations were carried out on NEC-SX9 at Osaka University and SR16000 at YITP in Kyoto University.
This study is supported in part by JSPS KAKENHI Grants Numbers JP25800170, JP25287046, JP26400281, JP15K17667,
and by MEXT as ``Priority Issue on Post-K computer'' (Elucidation of the Fundamental Laws and Evolution of the Universe)
and SPIRE (Strategic Program for Innovative REsearch).
T.H. was partly supported by RIKEN iTHES Project.
\end{acknowledgments}
%%%%%%%%%%%%%%%%%%%%%%%%%%%%%%%%%%%%%%%%%%%%%%%%%%

%%%%%%%%%%%%%%%%%%%%%%%%%%%%%%%%%%%%%%%%%%%%%%%%%%%%%%%%%%%%%%%%%%

\appendix

%%%%%%%%%%%%%%%%%%%%%%%%%%%%%%%%%%%%%%%%%%%%%%%%%%%%%%%%%%%%%%%%%%%%%%%%%%
\section{Supplemental Material}
%%%%%%%%%%%%%%%%%%%%%%%%%%%%%%%%%%%%%%%%%%%%%%%%%%%%%%%%%%%%%%%%%%%%%%%%%%

%%%%%%%%%%%%%%%%%%%%%%%%%%%%%%%%%%%%%%%%%%%%%%%%%%%%%%%%%%%%%%%%%%%%%%%%%%
\section{Complex poles of the coupled-channel $S$ matrix}
%%%%%%%%%%%%%%%%%%%%%%%%%%%%%%%%%%%%%%%%%%%%%%%%%%%%%%%%%%%%%%%%%%%%%%%%%%
%%%%%%%%%%%%%%%%%%%%%%%%%%%%%%%%%%%%%%%%%%%%%%%%%%%%%%%%%%%%%%%%%%%%%%%%%%
\begin{figure*}[bh]
\begin{center}
\includegraphics[width=0.39\textwidth,clip]{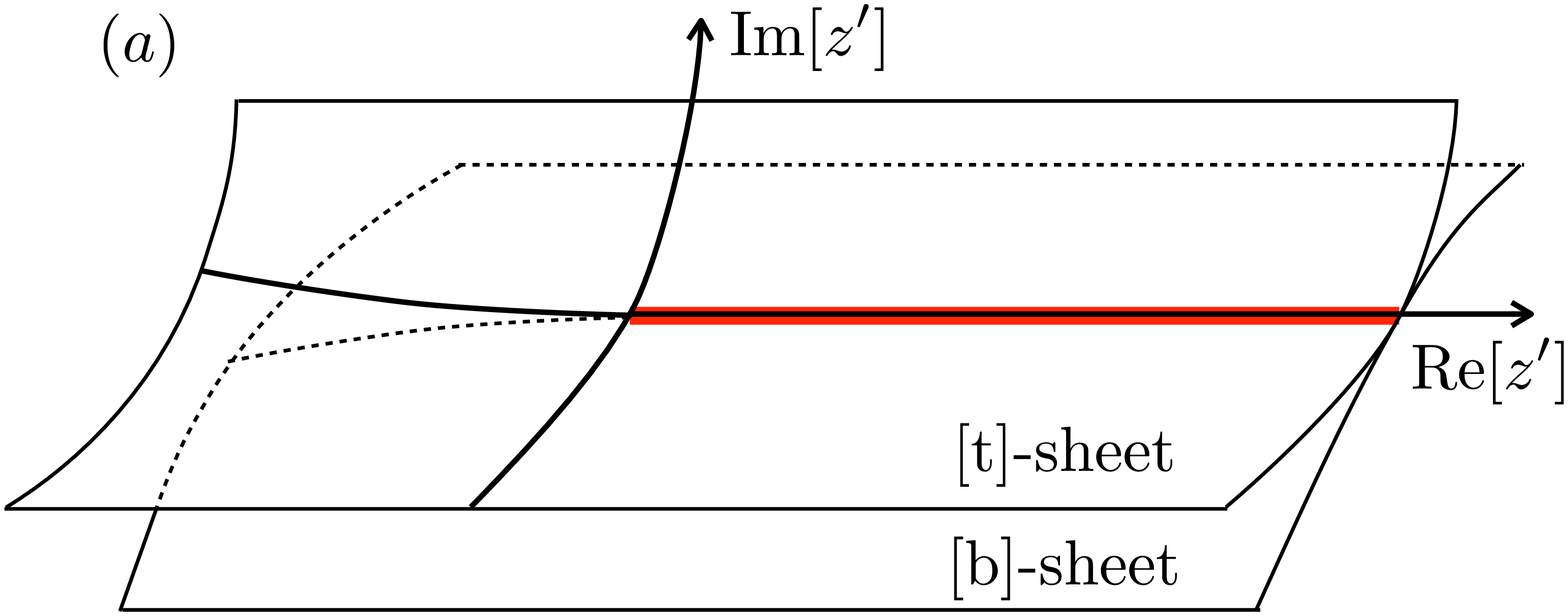}
%\\ \vspace{0.5cm}
\includegraphics[width=0.39\textwidth,clip]{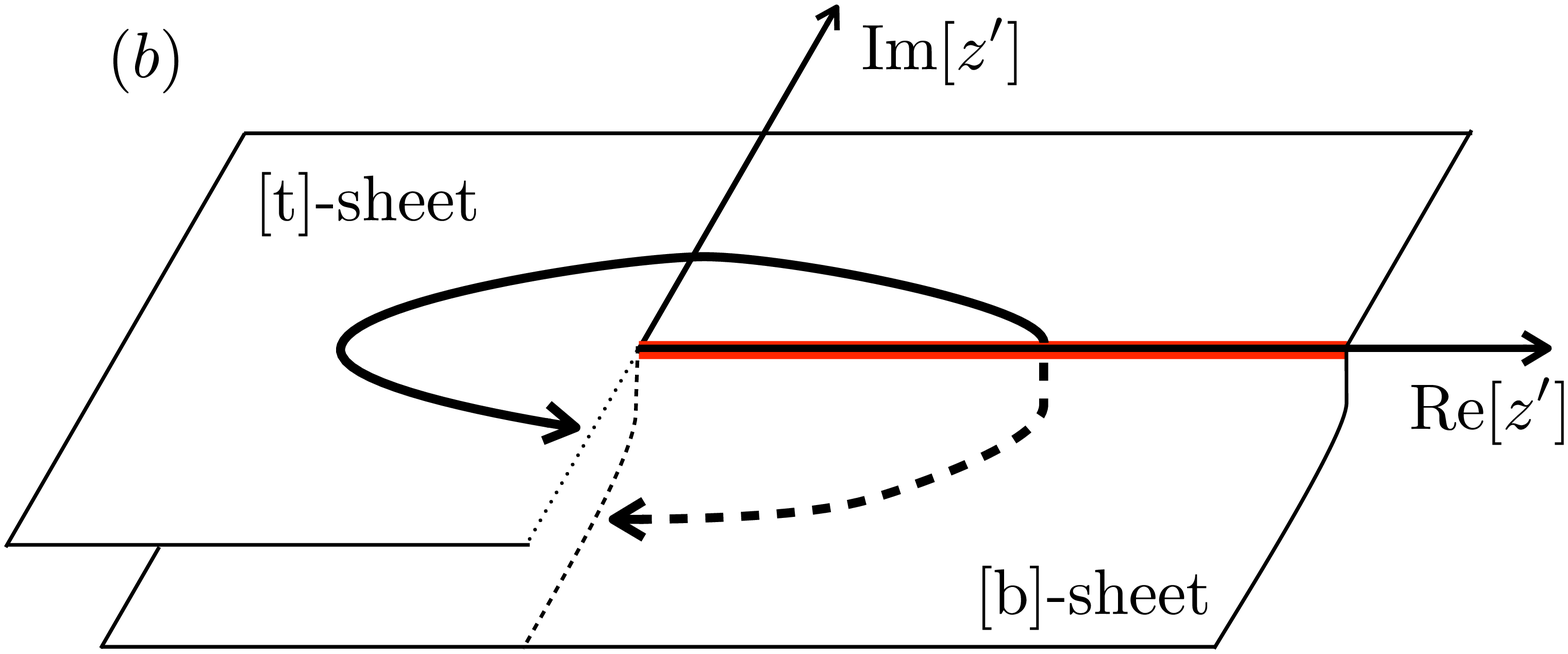}
%\\ \vspace{0.5cm}
\includegraphics[width=0.19\textwidth,clip]{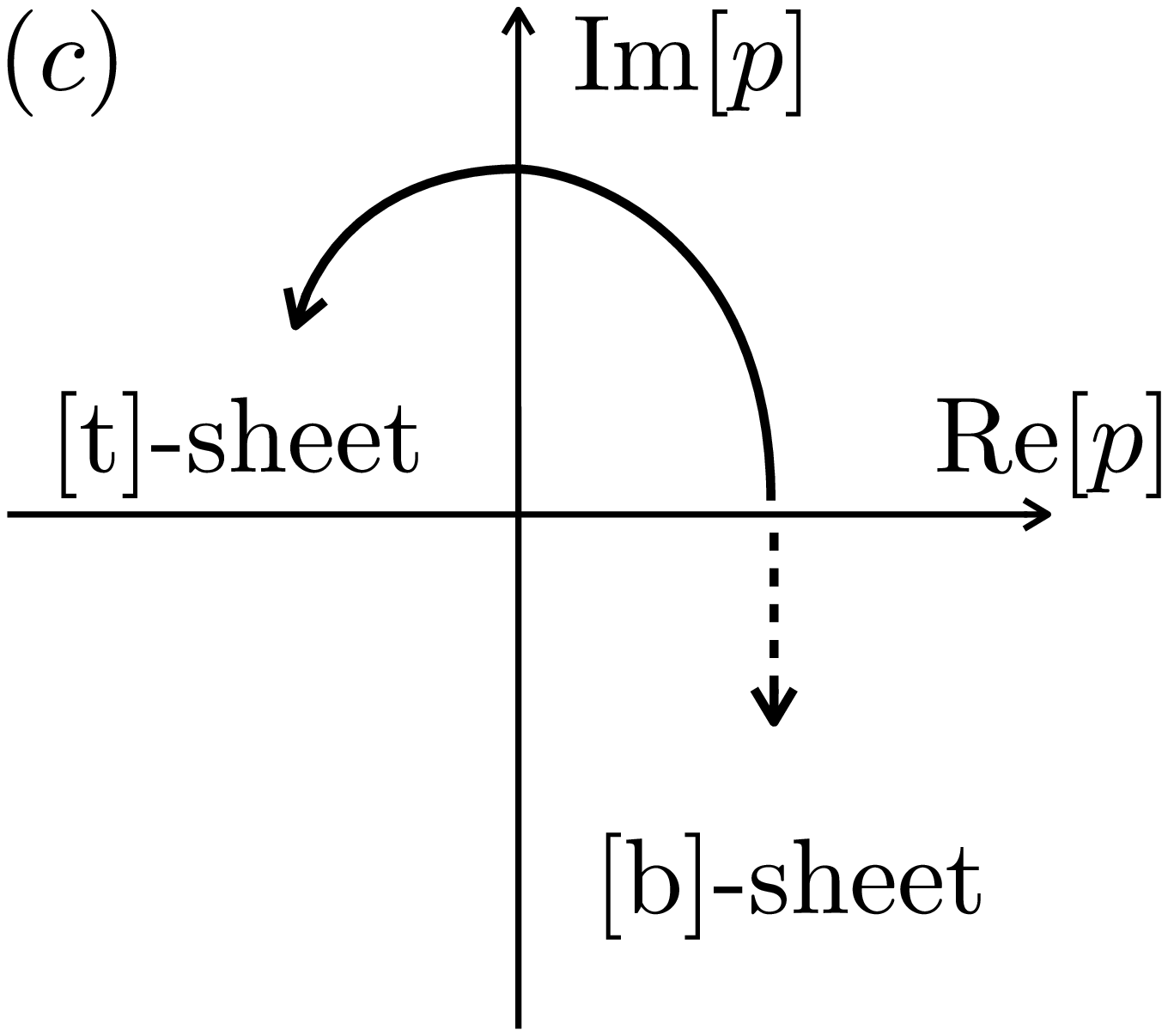}
\end{center}
\caption{
Structure of Riemann sheets for single-channel case with the notation of Ref.~\cite{Pearce:1988rk}. 
The complex energy $z$ is related to the complex momentum $p$ by $z \equiv m_1+m_2+p^2/(2\mu)$
with $\mu$ being the reduced mass.
In Fig.~\ref{fig:A1}(a) with $z' \equiv z- (m_1+m_2)$, 
the upper-half of the [$t$]-sheet is continuously connected to the lower-half [$b$]-sheet across the branch cut denoted by the red line.
In Fig.~\ref{fig:A1}(b), which is equivalent to Fig.~\ref{fig:A1}(a), 
will be used later for illustrating the Riemann sheets in the coupled-channel case. 
The [$t$]-sheet and the [$b$]-sheet correspond to 
the upper-half and the lower-half of the complex $p$-plane as shown in Fig.~\ref{fig:A1}(c). 
}
\label{fig:A1}
\end{figure*}
%%%%%%%%%%%%%%%%%%%%%%%%%%%%%%%%%%%%%%%%%%%%%%%%%%%%%%%%%%%%%%%%%%%%%%%%%%

%%%%%%%%%%%%%%%%%%%%%%%%%%%%%%%%%%%%%%%%%%%%%%%%%%%%%%%%%%%%%%%%%%%%%%%%%%
\begin{figure*}[b]
\begin{center}
\includegraphics[width=0.75\textwidth,clip]{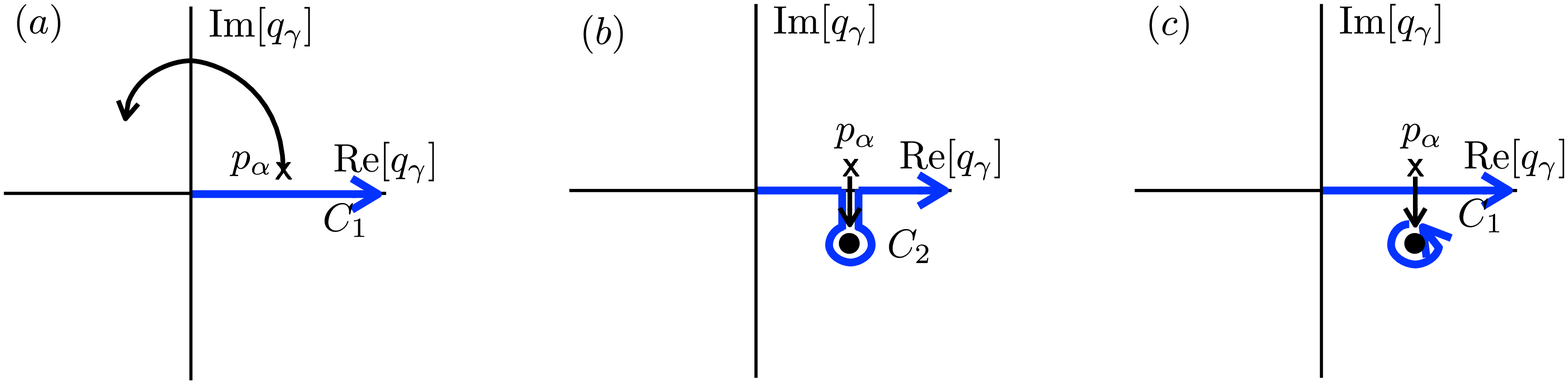}
\end{center}
\caption{
The integration contours in the complex momentum plane of $q_{\gamma}$.
The contour $C_{1}$ lies on the real axis in Figs.~\ref{fig:A2}(a) and \ref{fig:A2}(c), 
while the contour $C_{2}$ is deformed into the forth quadrant in the $q_{\gamma}$ plane 
due to the analytic continuation as shown in Fig.~\ref{fig:A2}(b).
We note that Figs.~\ref{fig:A2}(b) and \ref{fig:A2}(c) correspond to  Fig. 1(d) in Ref.~\cite{Pearce:1988rk}.
}
\label{fig:A2}
\end{figure*}
%%%%%%%%%%%%%%%%%%%%%%%%%%%%%%%%%%%%%%%%%%%%%%%%%%%%%%%%%%%%%%%%%%%%%%%%%%

Let us first review briefly the analytic continuation of two-body amplitudes in the complex energy plane
following Ref.~\cite{Pearce:1988rk}.
The coupled-channel Lippmann-Schwinger (LS) equation for the two-body $T$ matrix $t^{\alpha \beta}$ 
[$\alpha, \beta = \pi J/\psi, \rho \eta_c, \bar{D}D^{*}$] reads
%-------------------------------------------------------------------------
%\begin{widetext}
\begin{align}
& t^{\alpha \beta}(\vec{p}_{\alpha}, \vec{p}_{\beta}; W_{\mathrm{c.m.}})
= 
V^{\alpha \beta}(\vec{p}_{\alpha}, \vec{p}_{\beta}) \nonumber \\
& ~~~~
+ \sum_{\gamma} \int d\vec{q}_{\gamma}
\frac{ V^{\alpha \gamma}(\vec{p}_{\alpha}, \vec{q}_{\gamma}) 
t^{\gamma \beta}(\vec{q}_{\gamma}, \vec{p}_{\beta}; W_{\mathrm{c.m.}}) }
{ W_{\mathrm{c.m.}} - E_{\gamma}(\vec{q}_{\gamma}) + i \epsilon } ~,
\nonumber
%\label{eq:A.LS}
\end{align}
%\end{widetext}
%-------------------------------------------------------------------------
where $\vec{p}_{\alpha}$ ($\vec{q}_{\gamma}$) is 
the real on-shell (off-shell) momentum of the two-meson state in channel $\alpha$ ($\gamma$).
The scattering energy in the center-of-mass frame is
$W_{\mathrm{c.m.}} = m_1^{\alpha} + m_2^{\alpha} + \vec{p}^2_{\alpha}/2\mu^{\alpha}$, while 
$E_{\gamma}(\vec{q}_{\gamma}) = m_1^{\gamma} + m_2^{\gamma} + \vec{q}^2_{\gamma}/2\mu^{\gamma}$
represents the energy of the intermediate states in channel $\gamma$.
After the $s$-wave projection, the coupled-channel LS equation 
for the $s$-wave $T$ matrix $t^{\alpha \beta}_{(\ell =0)}$ can be written in terms of 
$p_{\alpha,\beta} = \left| \vec{p}_{\alpha,\beta} \right|$ 
and $q_{\gamma} = \left| \vec{q}_{\gamma} \right|$ as
%-------------------------------------------------------------------------
\begin{align}
& t^{\alpha \beta}_{(\ell=0)}(p_{\alpha}, p_{\beta}; W_{\mathrm{c.m.}})
=
V^{\alpha \beta}_{(\ell=0)}(p_{\alpha}, p_{\beta})  \nonumber \\
& ~~
+ \sum_{\gamma} \int d q_{\gamma} q_{\gamma}^{2}
\frac{ V^{\alpha \gamma}_{(\ell=0)}(p_{\alpha}, q_{\gamma}) 
t^{\gamma \beta}_{(\ell=0)}(q_{\gamma}, p_{\beta}; W_{\mathrm{c.m.}}) }
{ W_{\mathrm{c.m.}} - E_{\gamma}(q_{\gamma}) + i \epsilon } ~.
\label{eq:A.LS_s-wave}
\end{align}
%-------------------------------------------------------------------------
This is  related to the on-shell $S$ matrix as
%-------------------------------------------------------------------------
\begin{align}
&
S^{\alpha \beta}_{(\ell=0)}(p_{\alpha}, p_{\beta}; W_{\mathrm{c.m.}})
 =  \nonumber \\
& ~~~~
\delta_{\alpha \beta}
- 2 \pi i
\sqrt{ \mu^{\alpha} p_{\alpha}  \mu^{\beta} p_{\beta} }
\ t^{\alpha \beta}_{(\ell=0)}(p_{\alpha}, p_{\beta}; W_{\mathrm{c.m.}}) ~.
\nonumber
%\label{eq:A.S-mat_s-wave}
\end{align}
%-------------------------------------------------------------------------

To study the pole structure of the $Z_{c}(3900)$ in the complex energy plane,
we need to carry out  analytic continuation of Eq.~(\ref{eq:A.LS_s-wave}) 
in terms of  the complex energy $z \equiv m_1^{\alpha} + m_2^{\alpha} + p_{\alpha}^2/2\mu^{\alpha}$
with $p_{\alpha}$ being the complex momentum.
The top sheet and the bottom sheet 
(or the [$t$]-sheet and the [$b$]-sheet according to the notation of Ref.~\cite{Pearce:1988rk})
are joined along the branch cut on the real axis 
starting from $z' \equiv z - (m_1^{\alpha} + m_2^{\alpha}) =0 $ 
as shown in Fig.~\ref{fig:A1}(a). 
An alternative but equivalent way of illustrating the same structure is 
given by Fig.~\ref{fig:A1}(b) which is found to be more useful in multi-channel cases 
as we will see later.
Fig.~\ref{fig:A1}(c) shows the complex-momentum plane 
where the [$t$]-sheet and the [$b$]-sheet correspond to $0 \le \arg p_{\alpha} < \pi$ and $\pi \le \arg p_{\alpha} < 2\pi$, respectively.

We define the analytic continuation of the integral in Eq.~(\ref{eq:A.LS_s-wave}) as $I^{\gamma}(z)$.
For $z$ located in the [$t$]-sheet, the integral can be carried out 
by choosing  the contour $C_1$ for $q_\gamma$-integration (see Fig.~\ref{fig:A2}(a)):
%-------------------------------------------------------------------------
\begin{eqnarray}
I_{\rm [t]}^{\gamma}(z)
& = &
\int_{C_{1}} d q_{\gamma} q_{\gamma}^{2}
\frac{ V^{\alpha \gamma}_{(\ell=0)}(p_{\alpha}, q_{\gamma}) 
t^{\gamma \beta}_{(\ell=0)}(q_\gamma, p_{\beta}; z) }
{ z - E_{\gamma}(q_{\gamma}) } ~.
\nonumber
\end{eqnarray}
%-------------------------------------------------------------------------
On the other hand, for $z$ located in the [$b$]-sheet, the integration contour 
should be chosen to be $C_2$ for analytic continuation (see Fig.~\ref{fig:A2}(b)).
This is equivalent to picking up the anti-clockwise residue 
at the pole $+$ the integration along the contour $C_1$ 
as shown in Fig.~\ref{fig:A2}(c):
%-------------------------------------------------------------------------
\begin{eqnarray}
I_{\rm [b]}^{\gamma}(z)
& = &
\int_{C_{2}} d q_{\gamma} q_{\gamma}^{2}
\frac{ V^{\alpha \gamma}_{(\ell=0)}(p_{\alpha}, q_{\gamma}) 
t^{\gamma \beta}_{(\ell=0)}(q_{\gamma}, p_{\beta}; z) }
{ z - E_{\gamma}(q_{\gamma}) } 
\nonumber \\
& = &
\int_{C_{1}} d q_{\gamma} q_{\gamma}^{2}
\frac{ V^{\alpha \gamma}_{(\ell=0)}(p_{\alpha}, q_{\gamma}) 
t^{\gamma \beta}_{(\ell=0)}(q_{\gamma}, p_{\beta}; z) }
{ z - E_{\gamma}(q_{\gamma}) }   \nonumber \\
& &
- 2\pi i \mu^{\gamma} p_{\gamma} 
V^{\alpha \gamma}_{(\ell=0)}(p_{\alpha}, p_{\gamma})
t^{\gamma \beta}_{(\ell=0)}(p_{\gamma}, p_{\beta}; z) ~.
\nonumber
\end{eqnarray}
%-------------------------------------------------------------------------
Thus,
we obtain a coupled-channel LS equation 
defined on the complex energy plane \cite{Pearce:1988rk}:
%-------------------------------------------------------------------------
\begin{eqnarray}
t^{\alpha \beta}_{(\ell=0)}(p_{\alpha}, p_{\beta}; z)
& = &
V^{\alpha \beta}_{(\ell=0)}(p_{\alpha}, p_{\beta})
+ \sum_{\gamma} I_{\rm [t, b]}^{\gamma}(z) ~.
\label{eq:CCC}
\end{eqnarray}
%-------------------------------------------------------------------------

\begin{figure*}[b]
\begin{center}
\includegraphics[width=0.60\textwidth,clip]{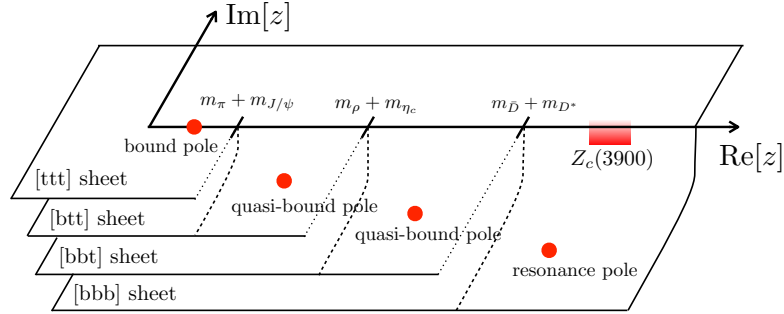}
\end{center}
\caption{
The complex energy plane in the
$\pi J/\psi$-$\rho \eta_c$-$\bar{D}D^{*}$ coupled-channel system.
The energy  relevant to $Z_c(3900)$ is indicated by shaded area, and
 examples of a possible resonance pole,  quasi-bound poles and a bound pole are 
 illustrated by the red filled circles.}
\label{fig:A3}
\end{figure*}

With Eq.~(\ref{eq:CCC}) for the $\pi J/\psi \text{-} \rho \eta_c \text{-} \bar{D}D^{*}$ system,
we now examine pole positions of the coupled-channel $S$ matrix which has 8 Riemann sheets originating from the existence of three thresholds.
These sheets are characterized by the notation [$xyz$]  with $x$, $y$ and $z$ taking either $t$ or $b$, according to Ref.~\cite{Pearce:1988rk}.
Among 8 sheets, the poles near the real axis on the [$bbb$], [$bbt$], [$btt$] and [$ttt$] sheets 
are most relevant for the scattering observables, since these sheets are directly
connected to the physical region.  On the other hand, the poles
on the [$tbt$], [$ttb$], [$tbb$] and [$btb$] sheets hardly affect the  scattering observables, since they
are not directly connected to the physical region.
Shown in Fig.~\ref{fig:A3} are some examples of ``possible'' poles on the [$bbb$], [$bbt$], [$btt$] and [$ttt$] sheets: 
 a $\bar{D}D^{*}$ resonance  pole near the real axis on the [$bbb$]-sheet,   
 a $\bar{D}D^{*}$ quasi-bound  pole  near the real axis on the [$bbt$]-sheet, 
 a $\rho \eta_{c}$  quasi-bound pole  near the real axis on the [$btt$]-sheet,
and a $\pi J/\psi$ bound  pole on the [$ttt$] sheet below the  threshold.
The energy region relevant to $Z_c(3900)$ is shown by the shaded area in Fig.~\ref{fig:A3}.

%%%%%%%%%%%%%%%%%%%%%%%%%%%%%%%%%%%%%%%%%%%%%%%%%%%%%%%%%%%%%%%%%%%%%%%%%%
%%%%%%%%%%%%%%%%%%%%%%%%%%%%%%%%%%%%%%%%%%%%%%%%%%%%%%%%%%%%%%%%%%%%%%%%%%

%%%%%%%%%%%%%%%%%%%%%%%%%%%%%%%%%%%%%%%%%%%%%%%%%%%%%%%%%%%%%%%%%%%%%%%%%%
\begin{figure*}[t]
\begin{center}
\includegraphics[width=0.40\textwidth,clip]{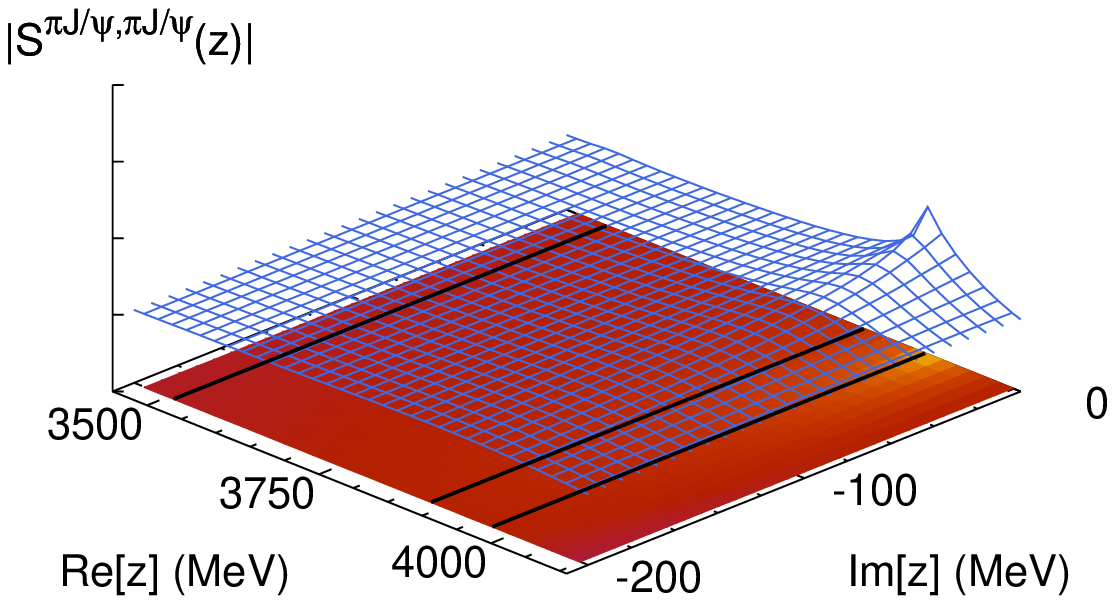}
\put(-180,120){(a)}
\hspace{1cm}
\includegraphics[width=0.40\textwidth,clip]{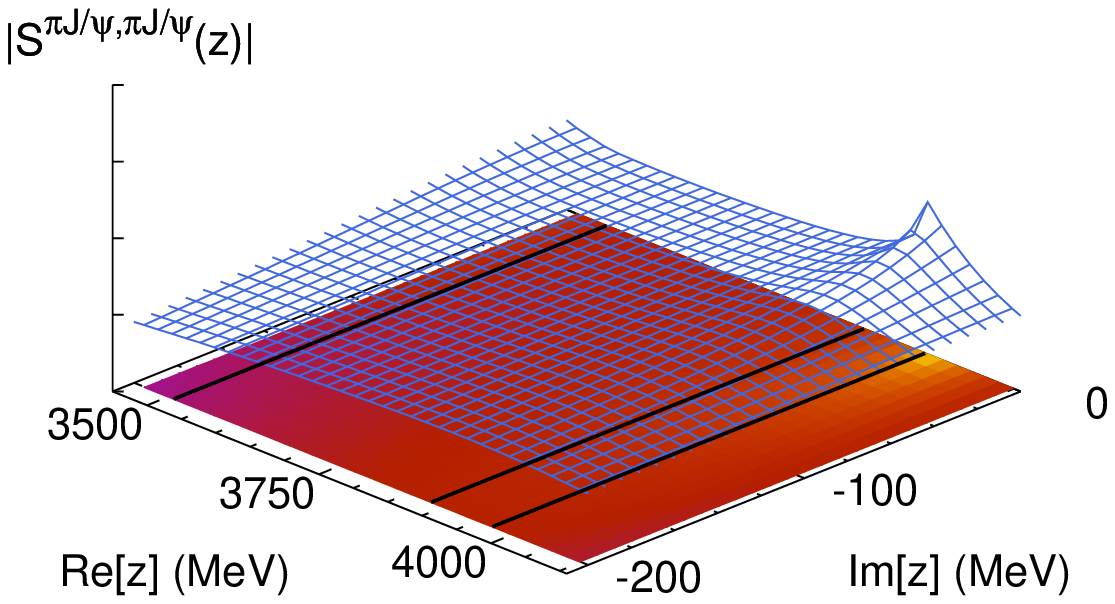}
\put(-180,120){(b)}\\
%\vspace{0.1cm}
\includegraphics[width=0.40\textwidth,clip]{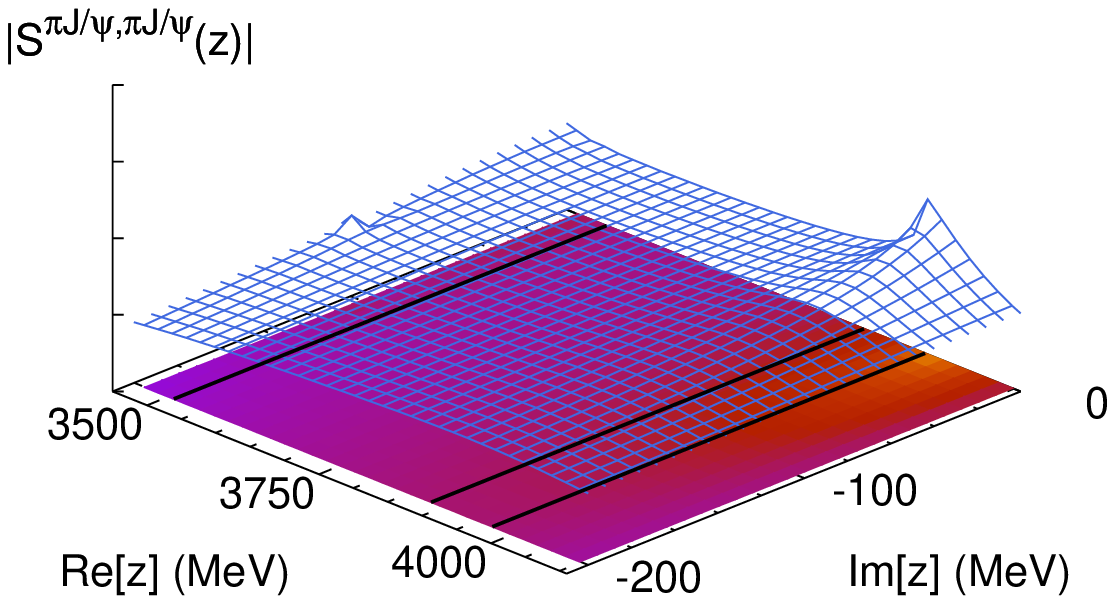}
\put(-180,120){(c)}
\hspace{1cm}
\includegraphics[width=0.40\textwidth,clip]{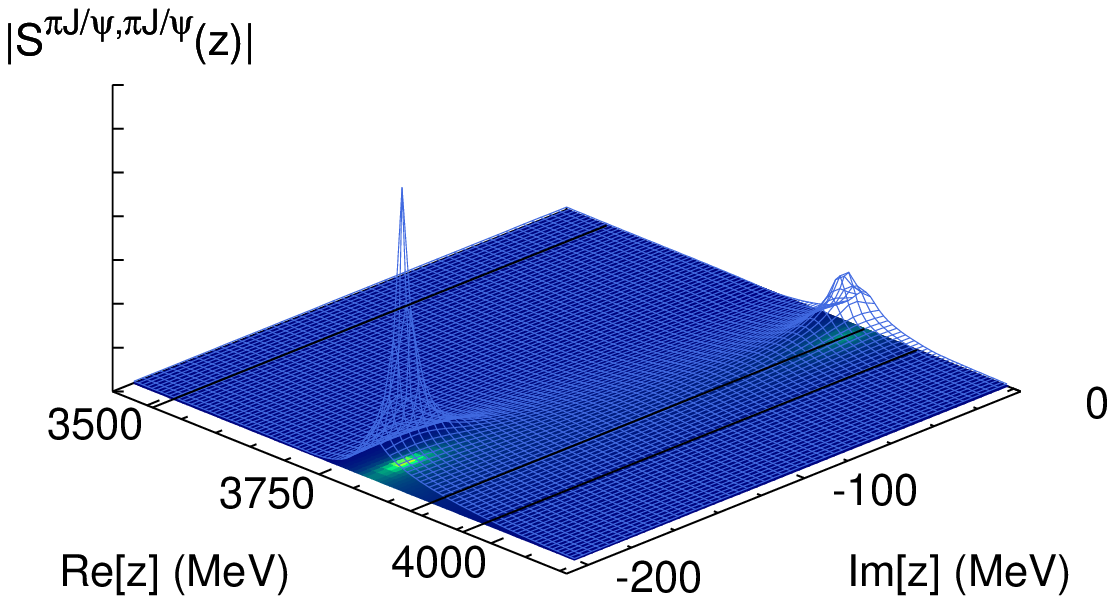}
\put(-180,120){(d)}
\end{center}
\caption{
The absolute magnitude of the $S$ matrix in the  $\pi J/\psi \text{-} \pi J/\psi$ channel on the 
(a) [$ttt$], (b) [$btt$], (c) [$bbt$] and (d) [$bbb$] sheets in the notation of Ref.~\cite{Pearce:1988rk}. 
The quark mass corresponds to case I of Table I in the main text.
}
\label{fig:A4}
\end{figure*}
%%%%%%%%%%%%%%%%%%%%%%%%%%%%%%%%%%%%%%%%%%%%%%%%%%%%%%%%%%%%%%%%%%%%%%%%%%
%%%%%%%%%%%%%%%%%%%%%%%%%%%%%%%%%%%%%%%%%%%%%%%%%%%%%%%%%%%%%%%%%%%%%%%%%%
\begin{figure*}[t]
\begin{center}
\includegraphics[width=0.40\textwidth,clip]{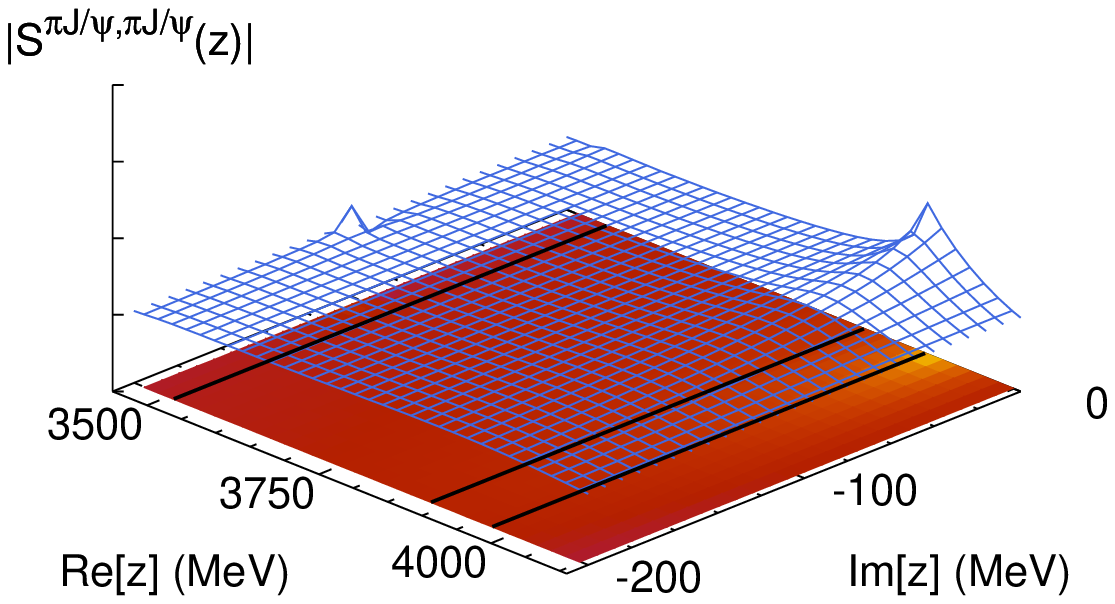}
\put(-180,120){(a)}
\hspace{1cm}
\includegraphics[width=0.40\textwidth,clip]{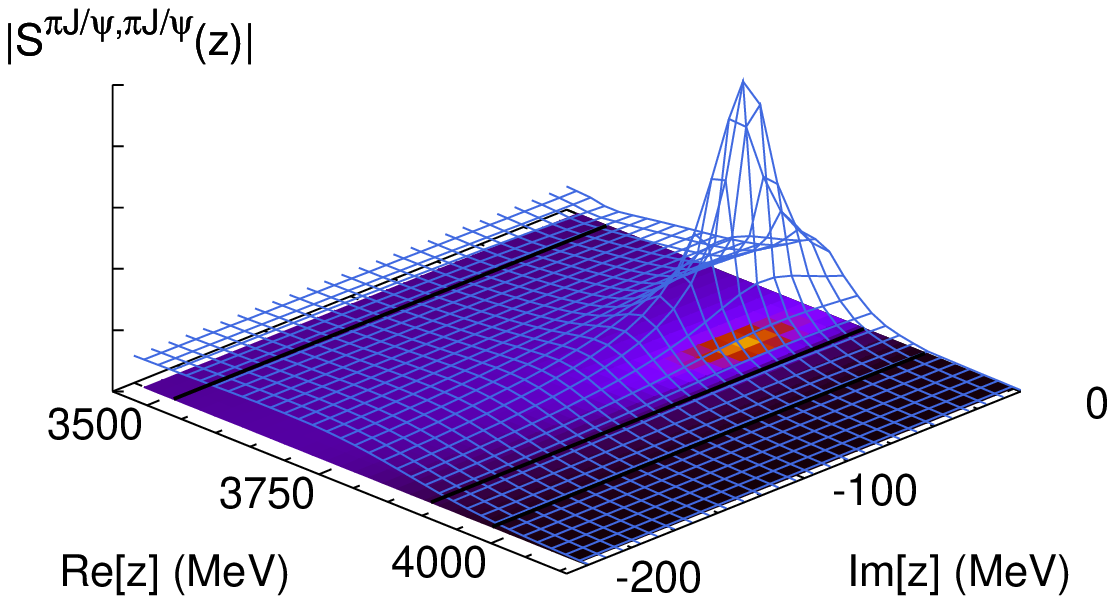}
\put(-180,120){(b)}\\
%\vspace{0.1cm}
\includegraphics[width=0.40\textwidth,clip]{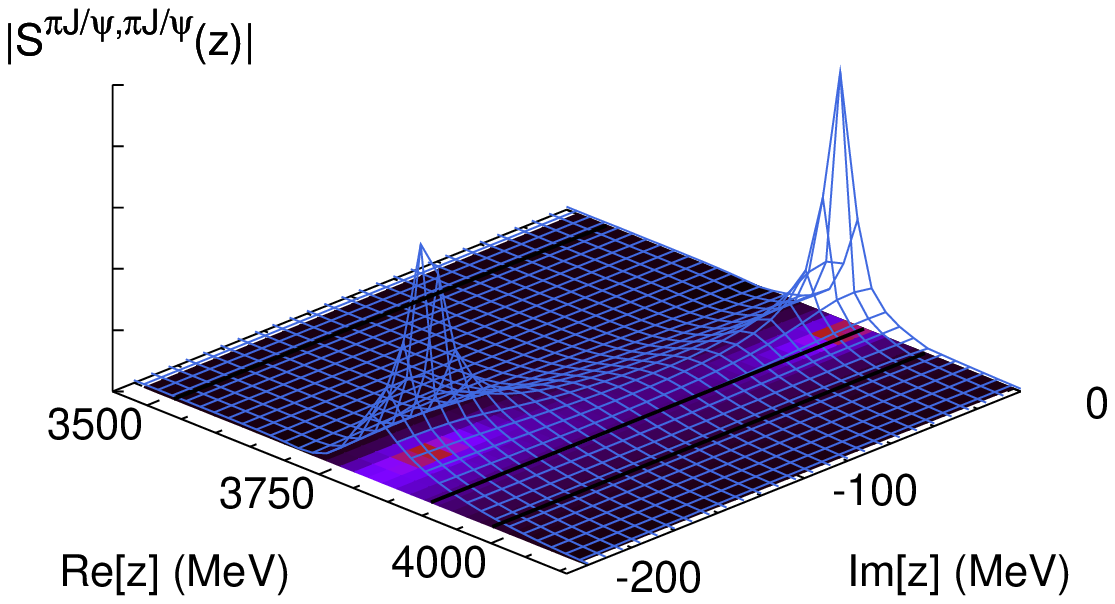}
\put(-180,120){(c)}
\hspace{1cm}
\includegraphics[width=0.40\textwidth,clip]{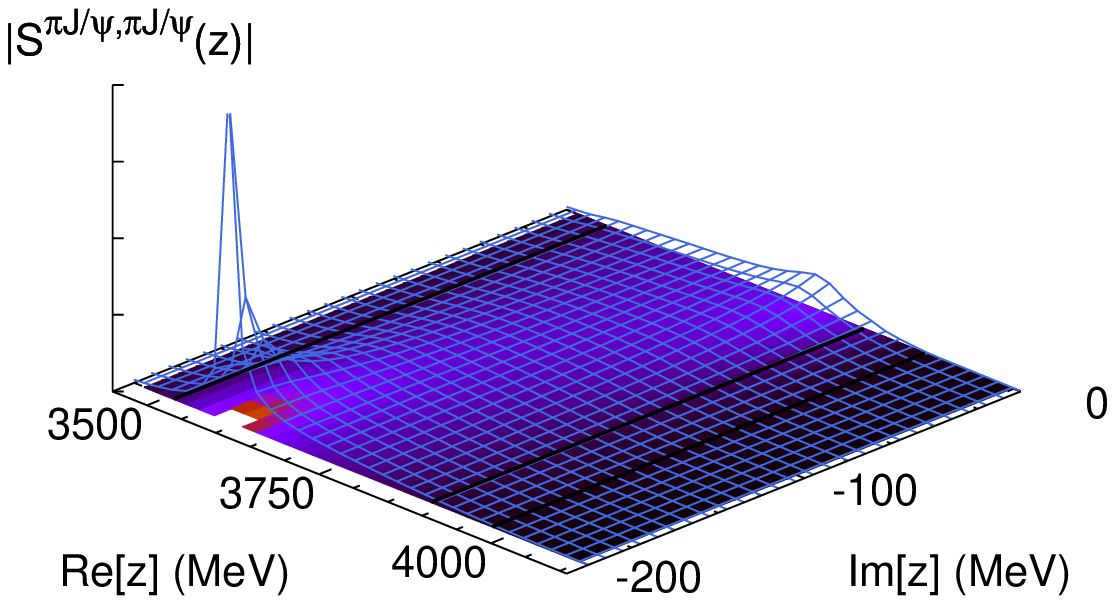}
\put(-180,120){(d)}
\end{center}
\caption{
Same as in Fig.~\ref{fig:A4}, but for the
(a) [$tbt$], (b) [$ttb$], (c) [$tbb$] and (d) [$btb$] sheets.
}
\label{fig:A6}
\end{figure*}
%%%%%%%%%%%%%%%%%%%%%%%%%%%%%%%%%%%%%%%%%%%%%%%%%%%%%%%%%%%%%%%%%%%%%%%%%%

\begin{figure*}[!h]
\begin{center}
\includegraphics[width=0.70\textwidth,clip]{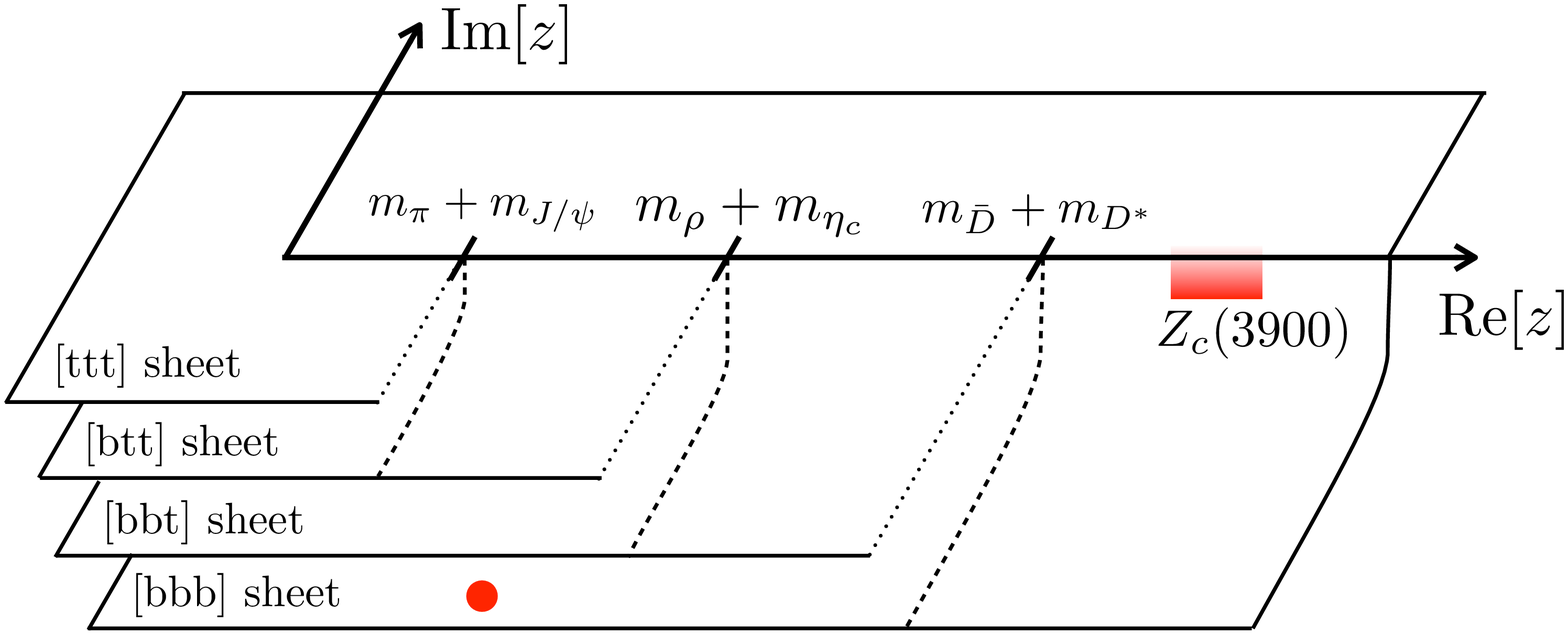}
\end{center}
\caption{
Illustration of a pole on the [$bbb$] sheet numerically found from the analytically continued $S$ matrix.
It is located far from the the physical region relevant for the  $Z_c(3900)$ indicated by the shaded area.
}
\label{fig:A5}
\end{figure*}

We have numerically searched through the poles on all 8 sheets in the $\pi J/\psi \text{-} \pi J/\psi$ $S$ matrix. 
In Fig.~\ref{fig:A4}, we show the absolute value of the $S$ matrix on (a) [$ttt$], (b) [$btt$], (c) [$bbt$] and (d) [$bbb$] sheets 
for case I in Table I of the main text.
On the [$ttt$], [$btt$] and [$bbt$] sheets, we do not find poles corresponding to the bound and quasi-bound states,
while, on the [$bbb$] sheet, we find a pole  with a large imaginary part,
$z_{\rm pole}-(m_{\bar{D}}+m_{D^{*}})=-167(94)(27)-i183(46)(19)$ MeV.
Here the mean value is calculated with the potential at $t=13$.
The first parenthesis indicates the statistical error from lattice data,
and the second parenthesis indicates the systematic error 
evaluated by the difference between the pole position at $t=13$ and that at $t=15$.
Since the pole is located far below the $\bar{D}D^{*}$ threshold, 
it does not affect the scattering observables.
Just for completeness, we show, in Fig.~\ref{fig:A6},  
the absolute value of the $S$ matrix on the (a) [$tbt$], (b) [$ttb$], (c) [$tbb$] and (d) [$btb$] sheets for case I of Table I in the main text.
We find poles on the [$ttb$], [$tbb$] and [$btb$] sheets, 
although they do not affect the observables as we mentioned before. 
The numerical results for the pole positions not only for case I but also for case II and III 
are summarized in Table~\ref{tab:A1}.
Also, the location of the pole on the [$bbb$] sheet is schematically illustrated in Fig.~\ref{fig:A5}.
  
With all these analyses of coupled-channel $S$ matrix,  
we conclude that the $Z_c(3900)$ is neither a conventional resonance nor a quasi-bound state.

%%%%%%%%%%%%%%%%%%%%%%%%%%%%%%%%%%%%%%%%%%%%%%%%%%%%%%%%%%%%%%%%%%%%%%%%%%
%%%%%%%%%%%%%%%%%%%%%%%%%%%%%%%%%%%%%%%%%%%%%%%%%%%%%%%%%%%%%%%%%%%%%%%%%%

%%%%%%%%%%%%%%%%%%%%%%%%%%%%%%%%%%%%%%%%%%%%%%%%%%%%%%%%%%%%%%%%%%%%%%%%%%
\begin{table*}[!h]
   \centering
   %\topcaption{Table captions are better up top} % requires the topcapt package
   \begin{tabular}{c|c|c|c|c}
      \hline
      \hline
 Case  &   [$bbb$]   & [$ttb$] & [$tbb$] & [$btb$] \\
      \hline 
I   & $-167(94)(27)-i183(46)(19)$ & $-146(112)(108) - i 38(148)(32)$ & $-177(116)(61) - i 175(30)(22)$ & $-369(129)(102) - i 207(61)(20)$  \\
 & & $-93(55)(21) - i 9(25)(7)$  & \\
      \hline 
 II  & $-128(76)(33)-i157(32)(19)$ & $-102(84)(45) - i 14(11)(7)$ & $-141(92)(64) - i 151(149)(132)$ & $-322(141)(111) - i 114(96)(75)$ \\
& & $-59(67)(11) - i 3(12)(1)$ & \\
      \hline 
 III & $-190(56)(42)-i44(27)(27)$  & $-100(48)(29) -i 7(37)(17)$ & $-127(52)(43) -i 199(44)(28)$ & $-356(108)(28) -i 277(138)(95)$ \\
& & $-53(30)(5) -i 2(11)(3)$ & \\
      \hline 
      \hline 
   \end{tabular}
   \caption{
   The pole positions $z'_{\rm pole}\equiv z_{\rm pole} - m_{\bar{D}} - m_{D^{*}}$ in MeV
   in different sheets with three cases for the pion mass.  Case I, II and III correspond to 
   those of Table I in the main text.
   The mean value is calculated with the potential at $t=13$.
   The first parenthesis indicates the statistical error from lattice data,
   and the second parenthesis indicates the systematic error 
   from the difference between the pole position at $t=13$ and that at $t=15$.
     }
   \label{tab:A1}
\end{table*}
%%%%%%%%%%%%%%%%%%%%%%%%%%%%%%%%%%%%%%%%%%%%%%%%%%%%%%%%%%%%%%%%%%%%%%%%%%

%%%%%%%%%%%%%%%%%%%%%%%%%%%%%%%%%%%%%%%%%%%%%%%%%%%%%%%%%%%%%%%%%%%%%%%%%%
\section{Derivation of three-body amplitude for Y(4260) decay}
\begin{figure*}[t]
\begin{center}
\includegraphics[width=0.85\textwidth,clip]{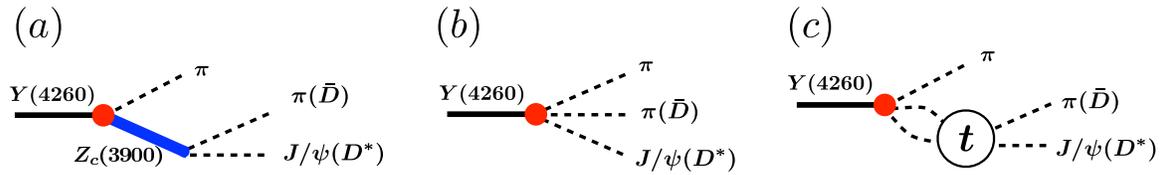}
\end{center}
\caption{
The diagrams of the $Y(4260) \to \pi \pi J/\psi (\pi \bar{D} D^{*})$ three-body decay.
(a) The background process and (b) the reaction process.
}
\label{fig:B1}
\end{figure*}

%%%%%%%%%%%%%%%%%%%%%%%%%%%%%%%%%%%%%%%%%%%%%%%%%%%%%%%%%%%%%%%%%%%%%%%%%%
Let us derive the three-body $T$ matrix~\cite{PDG} for the $Y(4260) \to \pi \pi J/\psi \ (\pi \bar{D} D^{*})$ decay.
We consider the $Y(4260)$ sequential decay {\it via} the $Z_c(3900)$ as shown in Fig.~\ref{fig:B1}(a).
The background process and the reaction process of the three-body decay are taken into account by 
Figs.~\ref{fig:B1}(b) and \ref{fig:B1}(c), respectively.

The three-body $T$ matrix for $Y(4260) \to \pi + \alpha$ ($\alpha = \pi J/\psi, ~ \bar{D} D^{*}$) in the $Y(4260)$ rest frame reads
%-----------------------------------------------------------------------
\begin{align}
& \langle Y(4260) | T^{Y \rightarrow \pi+\alpha}(W_3) |\vec{p}_{\alpha}, \vec{q}_{\alpha} \rangle
=    \nonumber \\
&
~~~~
\langle Y(4260) |
\left\{
\Gamma + \Gamma G_{0}(W_3) t( W_3, E_{\pi}(\vec{p}_{\alpha}) )
\right\}
|\vec{p}_{\alpha}, \vec{q}_{\alpha} \rangle ~,
\nonumber
\end{align}
%-----------------------------------------------------------------------
where $|Y(4260) \rangle$ denotes the $Y(4260)$ state,
and $|\vec{p}_{\alpha}, \vec{q}_{\alpha} \rangle$ is the three-body scattering state 
with spectator pion momentum $\vec{p}_{\alpha}$ and two-body subsystem with relative momentum $\vec{q}_{\alpha}$ in channel $\alpha$.
The relative momentum $\vec{q}_{\alpha}$ is defined in the two-body center-of-mass frame.
The three-body scattering state is normalized as
$\langle \vec{p}_{\alpha}, \vec{q}_{\alpha} | \vec{p}_{\beta}, \vec{q}_{\beta} \rangle = \delta_{\alpha \beta}
\delta(\vec{p}_{\alpha} - \vec{p}_{\beta}) \delta(\vec{q}_{\alpha}-\vec{q}_{\beta})$.
The mass of the $Y(4260)$ and the  energy of the spectator pion in channel $\alpha$
 are denoted by $W_3$ and $E_{\pi}(\vec{p}_{\alpha})$, respectively.
 The vertex operator of $Y(4260) \to \pi + \alpha$ is
 represented by $\Gamma$. 
Three-body free Green's function $G_0(W_3)$ and 
two-body $T$ matrix $t(W_3, E_{\pi}(\vec{p}_{\alpha}))$
are given by
%-----------------------------------------------------------------------
\begin{align}
& 
\langle \vec{p}_{\alpha}, \vec{q}_{\alpha} | G_{0}(W_3) | \vec{p}_{\beta}, \vec{q}_{\beta} \rangle
= 
\frac{
\delta_{\alpha \beta}
\delta(\vec{p}_{\alpha} - \vec{p}_{\beta}) \delta(\vec{q}_{\alpha}-\vec{q}_{\beta})
}
{W_3 - E_{\pi}(\vec{p}_{\alpha}) - E_{\alpha}(\vec{p}_{\alpha},\vec{q}_{\alpha}) + i \epsilon}  ~, \nonumber \\
&
\langle \vec{p}_{\alpha}, \vec{q}_{\alpha} | t(W_3, E_{\pi}(\vec{p}_{\alpha})) | \vec{p}_{\beta}, \vec{q}_{\beta} \rangle
\nonumber \\
& 
= 
\delta(\vec{p}_{\alpha} - \vec{p}_{\beta})
t^{\alpha \beta}(\vec{q}_{\alpha}, \vec{q}_{\beta}, \vec{p}_{\beta}; W_3)  \nonumber \\
&
=
\delta(\vec{p}_{\alpha} - \vec{p}_{\beta})
\biggl[
V^{\alpha \beta}(\vec{q}_{\alpha}, \vec{q}_{\beta}) 
\nonumber \\
&
~~~~~~~~~~~~
+ 
\sum_{\gamma} \int d\vec{q}_{\gamma}
\frac{
V^{\alpha \gamma}(\vec{q}_{\alpha}, \vec{q}_{\gamma})
t^{\gamma \beta}(\vec{q}_{\gamma}, \vec{q}_{\beta}, \vec{p}_{\beta}; W_3)
}{
W_3 - E_{\pi}(\vec{p}_{\gamma}) - E_{\gamma}(\vec{p}_{\gamma},\vec{q}_{\gamma}) + i\epsilon 
}
\biggr]  ~, \nonumber
\end{align}
%-----------------------------------------------------------------------
with
$E_{\alpha}(\vec{p}_{\alpha},\vec{q}_{\alpha}) = M_{\alpha}(\vec{q}_{\alpha}) + \vec{p}_{\alpha}^2 /2(m_1^{\alpha}+m_2^{\alpha})$
being the energy of the two-body subsystem,
where $M_{\alpha}(\vec{q}_{\alpha})$ is the two-body invariant mass in channel $\alpha$.

By modeling the primary vertex of $Y(4260) \to \pi + \alpha$ (red circle in Fig.~\ref{fig:B1})
by a complex constant, $C^{Y \to \pi + \alpha} = \langle Y(4260) | \Gamma | \vec{p}, \vec{q}_{\alpha} \rangle$,
we finally obtain the three-body $T$ matrix as
%-----------------------------------------------------------------------
\begin{align}
&
\langle Y(4260) | T^{Y \rightarrow \pi+\beta}(W_3) |\vec{p}_{\beta}, \vec{q}_{\beta} \rangle
\nonumber \\
&
= 
T^{Y\rightarrow \pi+\beta}(\vec{p}_{\beta}, \vec{q}_{\beta}; W_3)   \nonumber \\
& =
\sum_{\alpha}
C^{Y \rightarrow \pi+\alpha}  
\nonumber \\
& 
~~~~~~ \times 
\biggl(  \delta_{\alpha \beta} 
+ \int d\vec{q}_{\alpha} 
\frac{
t^{\alpha \beta}(\vec{q}_{\alpha}, \vec{q}_{\beta}, \vec{p}_{\beta}; W_3)
}{
W_3 - E_{\pi}(\vec{p}_{\beta}) - E_{\alpha}(\vec{p}_{\beta},\vec{q}_{\alpha}) + i\epsilon 
}
\biggr) ~.
\label{app:eq_decay}
\end{align}
%-----------------------------------------------------------------------

%%%%%%%%%%%%%%%%%%%%%%%%%%%%%%%%%%%%%%%%%%%%%%%%%%%%%%%%%%%%%%%%%%%%%%%%%%
%%%%%%%%%%%%%%%%%%%%%%%%%%%%%%%%%%%%%%%%%%%%%%%%%%%%%%%%%%%%%%%%%%%%%%%%%%

\end{document}